\documentclass[12pt]{iopart}

\usepackage{iopams}
\usepackage{graphicx}

\expandafter\let\csname equation*\endcsname\relax
\expandafter\let\csname endequation*\endcsname\relax
\usepackage{amssymb}
\usepackage{amsmath}
\usepackage{gensymb}

\usepackage{mathbbol}

\usepackage[colorlinks, linkcolor=blue, citecolor = blue, urlcolor = blue]{hyperref}
\usepackage{cite}

\usepackage{epsfig}
\usepackage{latexsym}
\usepackage{epic}
\usepackage{eepic}
\usepackage{psfrag}
\usepackage{rotating}
\usepackage{multirow,booktabs}
\usepackage{color}
\usepackage{cprotect} 
\usepackage{miller} 
\usepackage{xspace} 

\usepackage[export]{adjustbox}
\usepackage{graphicx}
 
\usepackage{algorithm}
\usepackage[]{algorithmicx,algpseudocode}

\usepackage{iopams}
\usepackage{stmaryrd} 
\usepackage{import}
\usepackage{breqn} 

\usepackage{array}

\usepackage[normalem]{ulem}

\renewcommand\vec{\mathbf}

\newcommand{\ddr}{\ensuremath{{\mathrm d}{\mathbf r} }\xspace} 
\newcommand{\rr}{\ensuremath{{\mathbf r} }\xspace}

\newcommand{\kk}{\ensuremath{{\mathbf k} }\xspace} 
\newcommand{\mm}{\ensuremath{{\mathbf m} }\xspace} 

\newcommand{\phii}[1][]{\ensuremath{\varphi}\xspace\ifx\relax#1\relax\else\ensuremath{\left(#1\right)}\xspace\fi}
 
\newcommand{\phiav}{\ensuremath{\bar{\phii}}\xspace} 
\newcommand{\BxO}{\ensuremath{B^x_0}\xspace} 
\newcommand{\DBO}{\ensuremath{\Delta B_0}\xspace}
\newcommand{\qo}{\ensuremath{q_0}\xspace} 

\newcommand{\FF}[1][]{\ensuremath{\mathcal{F}}\xspace\ifx\relax#1\relax\else\ensuremath{\left[#1\right]}\xspace\fi}
\newcommand{\FFm}[1][]{\ensuremath{\mathcal{F_{\mathrm m}}}\xspace\ifx\relax#1\relax\else\ensuremath{\left[#1\right]}\xspace\fi}
\newcommand{\FFpfc}[1][]{\ensuremath{\mathcal{F_{\mathrm PFC}}}\xspace\ifx\relax#1\relax\else\ensuremath{\left[#1\right]}\xspace\fi}
\newcommand{\FFcoup}[1][]{\ensuremath{\mathcal{F_{\mathrm c}}}\xspace\ifx\relax#1\relax\else\ensuremath{\left[#1\right]}\xspace\fi}

\newcommand{\ff}[1][]{\ensuremath{f}\ifx\relax#1\relax\else\ensuremath{\left(#1\right)}\fi}

\newcommand{\ffpfc}[1][]{\ensuremath{\ff_{\mathrm PFC}}\ifx\relax#1\relax\else\ensuremath{\left(#1\right)}\fi}
\newcommand{\ffid}[1][]{\ensuremath{\ff_{\mathrm id}}\ifx\relax#1\relax\else\ensuremath{\left(#1\right)}\fi}
\newcommand{\ffex}[1][]{\ensuremath{\ff_{\mathrm ex}}\ifx\relax#1\relax\else\ensuremath{\left(#1\right)}\fi}
\newcommand{\ffm}[1][]{\ensuremath{\ff_{\mathrm m}}\ifx\relax#1\relax\else\ensuremath{\left(#1\right)}\fi}
\newcommand{\ffcoup}[1][]{\ensuremath{\ff_{\mathrm c}}\ifx\relax#1\relax\else\ensuremath{\left(#1\right)}\fi}

\newcommand{\NN}[1][]{\ensuremath{\mathcal{N}}\ifx\relax#1\relax\else\ensuremath{\left[{#1}\right]}\fi}
\newcommand{\NNj}[1][]{\ensuremath{\mathcal{N}_j}\ifx\relax#1\relax\else\ensuremath{\left[{#1}\right]}\fi}
\newcommand{\NNjp}[1][]{\ensuremath{\mathcal{N}_j^+}\ifx\relax#1\relax\else\ensuremath{\left[{#1}\right]}\fi}
\newcommand{\NNjm}[1][]{\ensuremath{\mathcal{N}_j^-}\ifx\relax#1\relax\else\ensuremath{\left[{#1}\right]}\fi}

\newcommand{\LL}[1][]{\ensuremath{\mathcal{L}}\ifx\relax#1\relax\else\ensuremath{#1}\fi}
\newcommand{\LLj}[1][]{\ensuremath{\mathcal{L}_j}\ifx\relax#1\relax\else\ensuremath{#1}\fi}
\newcommand{\GGj}[1][]{\ensuremath{\mathcal{G}_j}\ifx\relax#1\relax\else\ensuremath{#1}\fi}

\newcommand{\np}[1][]{\ensuremath{#1}\ifx\relax#1\relax\else\ensuremath{^{n+1}}\fi}
\newcommand{\n}[1][]{\ensuremath{#1}\ifx\relax#1\relax\else\ensuremath{^{n}}\fi}

\newcommand \be {\begin{eqnarray}}
\newcommand \ee {\end{eqnarray}}

\newcommand{\kkj}{\ensuremath{\vec{k}_j}\xspace}
\newcommand{\DD}{\ensuremath{\vec{D}}\xspace}
\newcommand{\DDk}{\ensuremath{\vec{D}_{\mathrm k}}\xspace}
\newcommand{\RR}{\ensuremath{\vec{R}}\xspace}
\newcommand{\sof}[1][]{\ensuremath{\{#1\}}\xspace}
\newcommand{\sofk}{\ensuremath{\sof[\kkj]}\xspace}

\newcommand{\sofkdef}{\ensuremath{\sof[\kkj']}\xspace}
\newcommand{\Aj}{\ensuremath{A_j}\xspace}
\newcommand{\A}{\ensuremath{A}\xspace}
\newcommand{\Asq}{\ensuremath{A^2}\xspace}
\newcommand{\Ajcc}{\ensuremath{A^*_j}\xspace}

\newcommand{\sofA}{\ensuremath{\sof[\Aj]}\xspace}
\newcommand{\sofAcc}{\ensuremath{\sof[\Ajcc]}\xspace}

\newcommand{\meS}[1][]{\ensuremath{\mathrm S}\ifx\relax#1\relax\else\ensuremath{_{#1}}\fi}

\newcommand{\MM}[1][]{\ensuremath{\mathcal{M}}\ifx\relax#1\relax\else\ensuremath{#1}\fi}
\newcommand{\MMj}[1][]{\ensuremath{\mathcal{M}_j}\ifx\relax#1\relax\else\ensuremath{#1}\fi}

\newcommand{\ii}{\mathbb{i}}

\newcommand{\aalpha}{\ensuremath{\alpha_2}\xspace}
\newcommand{\arot}{\ensuremath{\theta}}

\newcommand*\colvec[3][]{
    \begin{pmatrix}\ifx\relax#1\relax\else#1\\\fi#2\\#3\end{pmatrix}
}

\begin{document}

\title[]{Magnetic APFC modeling and the influence of magneto-structural interactions on grain shrinkage}

\author{Rainer Backofen}
\address{Institute of Scientific Computing, Technische Universit\"at Dresden, 01062 Dresden, Germany}
\ead{rainer.backofen@tu-dresden.de}
\author{Marco Salvalaglio}
\address{Institute of Scientific Computing, Technische Universit\"at Dresden, 01062 Dresden, Germany}
\address{Dresden Centre for Computational Materials Science (DCMS), TU Dresden, 01062 Dresden, Germany}
\ead{marco.salvalaglio@tu-dresden.de}
\author{Axel Voigt}
\address{Institute of Scientific Computing, Technische Universit\"at Dresden, 01062 Dresden, Germany}
\address{Dresden Centre for Computational Materials Science (DCMS), TU Dresden, 01062 Dresden, Germany}
\ead{axel.voigt@tu-dresden.de}

\begin{abstract}
We derive the amplitude expansion for a phase-field-crystal (APFC) model that captures the basic physics of magneto-structural interactions. The symmetry breaking due to magnetization is demonstrated, and the characterization of the magnetic anisotropy for a BCC crystal is provided. 
This model enables a convenient coarse-grained description of crystalline structures, in particular when considering the features of the APFC model combined with numerical methods featuring inhomogeneous spatial resolution. This is shown by addressing the shrinkage of a spherical grain within a matrix, chosen as a prototypical system to demonstrate the influence of different magnetizations. These simulations serve as a proof of concept for the modeling of manipulation of dislocation networks and microstructures in ferromagnetic materials within the APFC model.
\end{abstract}

%
%
%
%
%
\section{Introduction}

External magnetic fields offer the possibility to manipulate microstructure in ferromagnetic materials. This has several potential applications in microstructural engineering \cite{Guillonetal_MT_2018}. However, a detailed understanding of the interactions between magnetic fields and solid-state matter transport is unavailable. This is due to the lack of a theoretical model which allows describing the fundamental physics of magneto-structural interactions in a multiscale approach, combining the dynamics of defects, dislocation networks, and grain boundaries with experimentally accessible microstructure evolution on diffusive time scales.

A promising approach in this direction is provided by a recently introduced phase-field-crystal (PFC) model that captures the basic physics of magnetocrystalline interactions \cite{FPK13,SSP15}. The PFC approach \cite{EKG02,EG04} is a continuum method that has been shown in numerous publications to capture the essential physics of atomic-scale elastic and plastic effects that accompany diffusive phase transformations, such as solidification, dislocation kinetics, and solid-state precipitation, see \cite{Emmerichetal_AP_2012} for a review. In \cite{FPK13} the PFC density is coupled with magnetization to generate a ferromagnetic solid below a critical temperature (the Curie temperature). The latter can depend on the local mean density. In \cite{SSP15} this PFC model is extended to multiferroic binary solid solutions and used to demonstrate the influence of magnetic fields on the growth of crystalline grains. The model, which is a system of evolution equations for the rescaled particle density field $\phii$ and an averaged magnetization $\mathbf{m}$, is used in \cite{BEV19} together with efficient and scalable numerical algorithms \cite{Elsayetal_ESAIM_2013} to study the role played by external magnetic fields on the evolution of defect structures and grain boundaries. The induced magnetic anisotropy has consequences also on larger scales.
They result from the dependence of the bulk energy of single crystals on the direction of the magnetization w.r.t. to crystallographic orientations.
Indeed, simulations show that the application of external magnetic fields affects the growth and coarsening of grains in microstructures, leading to the selection of preferential orientations. These results are in qualitative agreement with experiments,
e.g., on Zn and Ti sheets \cite{Molodovetal_SM_2006}, and classical grain growth simulations of Mullins type with an analytical magnetic driving force \cite{Barrales-Moraetal_CMS_2007}. The additional driving force, due to the
external magnetic field, also enhances the coarsening process in the simulations  \cite{BEV19}. This effect has been observed
experimentally, e.g., during annealing of FeCo under high steady magnetic fields \cite{Rivoirard_JOM_2013}. It has also been shown that the grain boundary mobility is anisotropic with respect to the applied magnetic field. This kinetic effect leads to elongated grains. The simulations show that under the influence of a strong external magnetic field, the magnetization $\mathbf{m}$ can be assumed to be constant. This simplifies the considered model and allows the simulation of larger systems, thus enabling the analysis of long-time scaling behaviors and various geometrical and topological properties in grain growth. In \cite{BV20} it is investigated how the scaling and characteristic geometric and topological properties change under the influence of strong magnetic fields. The results are in qualitative agreement with experiments for thin Zr sheets annealed under the influence of magnetic fields \cite{MB10}. However, even with these fundamental contributions, simulation-driven applications in microstructural engineering are still out of reach. While the microscopic details are well resolved with the considered PFC model and also experimental relevant time scales can be reached, the required spatial resolution essentially restricts physically relevant simulation to two-dimensional settings or relatively small samples in three dimensions, see, e.g., \cite{BGE06,Yamanaka2017}. 

Required for a true multiscale approach is a coarse-graining in space. The complex amplitude PFC (APFC) model originally introduced in \cite{Goldenfeld2005,Athreya2006} provides such a framework. The idea of the APFC approach is to model the amplitude of the density fluctuations instead of the density itself. This enables to reach larger spatial scales by still retaining essential microscopic effects, thus enabling mesoscale investigations of crystalline systems \cite{Huang2008,elder2017striped,SVE19,SalvalaglioPRL2021,Jreidini2021}. For a recent review of APFC models, we refer to \cite{Salvalaglio2022}. 

We here extend the magnetic PFC model \cite{FPK13,SSP15}, or to be more precise, the simplified version used in \cite{BV20}, to the APFC framework. In Section~\ref{sec::cntrAPFC} we formulate and explain the model. The symmetry breaking due to magnetization is discussed in Section~\ref{sec::symm}. In Section~\ref{sec::num} we report the main concept behind the numerical approach exploited to solve the coupled system of equations of the magnetic APFC model, and in Section~\ref{sec::res} we demonstrate the applicability for a simple three-dimensional setting. We consider a spherical grain in a matrix and simulate its evolution under the influence of a strong magnetic field. The results are compared with PFC and APFC simulations without magnetic field \cite{Yamanaka2017,Salvalaglio2018}. Finally we draw conclusions in Section~\ref{sec::con}.

\section{From magnetic PFC to magnetic APFC models}
\label{sec::cntrAPFC}

We consider here the limit of strong external magnetic fields. In this limit, the local magnetization is assumed to be constant in the crystal. The direction and the magnitude of the magnetization is solely defined by the external magnetic field. Then the magnetic PFC model in \cite{FPK13,SSP15} in terms of its free energy functional, reduces to:
\begin{equation}
  \FF[\phii]=\int_{\Omega} \left[ \frac{B^x_0}{2} \phii(\qo^2+\nabla^2)^2\phii
  +\frac{\Delta B_0}{2}\phii^2
  -\frac{\tau}{3}\phii^3+\frac{v}{4}\phii^4 + \frac{\aalpha}{2} \phii (\mathbf{m} \nabla)^2 \phii\right] \ddr,
    \label{eq::cntrPFCenergy}
\end{equation}
where $\phii$ denotes the scaled particle density and $\mathbf{m}$ the magnetization. $\qo$ defines the lattice spacing at equilibrium, which we set here to 1 without loss of generality. $\Omega$ is the domain of integration. $\BxO$, $\DBO$, $\tau$ and $v$ are parameters as introduced in reference~\cite{Elder2007}. Together with the average density $\phiav$ they define crystal structure and physical properties. Parametric studies and phase diagrams obtained by varying such parameters in classical formulations of the PFC model can be found in the literature \cite{EG04,Elder2007,TTG10,JA10b}. $\aalpha$ controls the strength of the magnetic interaction. The considered form $\aalpha \phii (\mm \nabla)^2 \phii$ provides the simplest possible term leading to symmetry breaking by the magnetization, \mm, which is assumed to be constant. The magnetization is scaled to unit length, $|\mathbf{m}| \,= 1$ without loss of generality. Then, the evolution equation for conserved dynamics reads
\begin{equation}
  \partial_t \phii = \nabla^2 \frac{\delta \FF}{\delta \phii},   \label{eq::PFC}
\end{equation}
with 
\begin{equation}
\frac{\delta \FF}{\delta \phii} =  B^x_0 (1+\nabla^2)^2\phii
  +\Delta B_0 \phii
  -\tau \phii^2+v\phii^3 + \aalpha (\mathbf{m} \nabla)^2 \phii.
  \end{equation}
  Note that this consists of a 6th order partial differential equation. 
  We chose model parameters such to describe a BCC crystal at equilibrium, similarly to references~\cite{EHP10,SBE17}. They are reported in Table~\ref{tab:paramAPFC}. The magnetic coupling is controlled by $\aalpha$, which is specified below. 
\begin{table}[hb]
\centering
\begin{tabular}{|c|c|c|c|c|}
\hline
  $\tau$  & $v$ & $\bar{\phii}$ & $B^x_0$ & $\Delta B_0$ \\
  \hline
  1 & 1 & 0 & 0.98 & 0.02 \\
  \hline
\end{tabular}
\caption{\label{tab:paramAPFC} Modeling parameters for magnetic PFC and APFC model~\cite{EHP10,SBE17}.}
\end{table}

In the APFC model the particle density $\phii$ is expressed in terms of the Fourier modes of a relaxed reference crystal with complex amplitudes $A_j(\mathbf{r})$ accounting for its deformation,
\begin{equation}
 \phii(\rr) = \phiav+\sum_{j=1}^{N}\left[ \Aj(\rr) {\rm e}^{\ii \kkj \cdot \rr} +  \Aj^*(\rr) {\rm e}^{-\ii \kkj \cdot \rr} \right],
\label{eq::APFCref}
\end{equation}
where  $\Aj^*(\rr)$ is the complex conjugate of $\Aj(\rr)$ in order to describe the density wave.
The reference crystal is then defined by a set of N reciprocal space vectors \kkj. The set \sofk explicitly entering the APFC model is typically chosen by considering the minimum number of modes describing the crystal, i.e. the shortest \kkj, with $-$\kkj accounted for through the complex conjugate in eq.~\eqref{eq::APFCref} \cite{Salvalaglio2022}, see also Figure~\ref{fig::kjsBCC}. A single crystal, corresponding to the reference structure without deformations, is then described by real and constant amplitudes. 
For deformed crystals, the set of reciprocal-space vectors, \sofk, is modified by a constant deformation \DD: $\sofk \rightarrow \sofkdef=\sof[\kkj \DD^{-1}]$. With respect to the original description, eq.~\eqref{eq::APFCref}, this leads to  $\Aj(\rr) \rightarrow \Aj(\rr) {\rm e}^{\ii \kkj (\DD^{-1}-1) \rr}$.  
Thus, a constant deformation leads to spatial varying complex amplitudes. The gradient of the phases of the amplitudes, $\kkj (\DD^{-1}-1)$, define their periodicity. The same holds also for rotations. 
Further on, the explicit space dependence for $\phii(\rr)$ and $\Aj(\rr)$ will be omitted.

   
To derive the corresponding magnetic APFC model, eq.~\eqref{eq::APFCref} is substituted in the PFC energy including the effect of magnetization, eq.~\eqref{eq::cntrPFCenergy}. If the amplitudes, \Aj, vary on larger length scales than the density wave, \phii, then the fluctuations on small scales can be averaged \cite{OSP13}. This can be formally done by multi-scale analysis \cite{EHP10, AGD06}, or renormalization techniques \cite{AGD06, OS12}.
All techniques show that only resonant terms contribute to the energy in this limit. That is, only terms, where the phases of the reference density wave, $\kkj \cdot \rr$,  cancel each other, remain: e.g. $\Aj \Ajcc = \Aj {\rm e}^{\ii \kkj \cdot \rr} \Ajcc {\rm e}^{-\ii \kkj \cdot \rr}$ are the only quadratic resonant terms. Following these concepts, the magnetic APFC energy reads
\begin{equation}
  \FF[\sofA]=\int_{\Omega} 
                    \sum_{j=1}^N  \left( B_0^x \Ajcc \GGj^2 \Aj
                      + \aalpha \Ajcc \MMj^2 \Aj \right)+ g^{\rm S}(\sofA) \ddr,
                    \label{eq::APFCenergy}
\end{equation}
with
\begin{equation}
g^{\rm S}(\sofA) = \sum_{j=1}^N  \left( -\frac{3v}{2}|\Aj|^4 \right)
                 + \frac{\Delta B_0}{2}A^2
                 + \frac{3v}{4}A^4
                 +f^{\rm S}(\sofA), \nonumber 
\end{equation}
and $A^2\equiv 2\sum_{j=1}^N |\Aj|^2$, $A^4 \equiv (A^2)^2$, $\GGj \equiv 1-|\kkj|^2+\nabla^2+2\ii\kkj \cdot \nabla$ and $\MMj=(\mm \cdot \nabla + \ii \mm \cdot \kkj) $.
A convenient choice of the reference crystal is the one corresponding to an expansion as in eq.~\eqref{eq::APFCref} which minimizes the free energy, eq.~\eqref{eq::APFCenergy}.
It corresponds to $|\kkj|=1$ with constant and real amplitude, whose values are dependent on model parameters \cite{Salvalaglio2022}. 
$f^{\rm S}$ is a polynomial function in  \sofA and \sofAcc. It depends on the reference crystal structure, see \cite{EHP10,SBE17,Salvalaglio2022}.
An example of $f^{\rm S}$ is given in Section~3 for a BCC crystal.
The energy reported above, eq.~\eqref{eq::APFCenergy}, can be split in three characteristic parts: 
\begin{itemize}
\item[i)] $\sum_{j=1}^N  B_0^x \Ajcc \mathcal{G}_j^2 \Aj$, encodes the dependence of the energy on amplitudes' phases through their gradients, and thus on lattice deformation with respect to the reference crystal. As such, it leads to the elastic properties of the model \cite{EG04,SVE19}. 
\item[ii)] $g^{\rm S}(\sofA)$, is a polynomial in \sofA, which is independent on phase of amplitudes. Thus, it does not change with deformation of the crystal. It contains the lattice dependent polynomial $f^{\rm S}(\sofA)$, and a term $\sum_{j=1}^N  \left( -\frac{3v}{2}|\Aj|^4 \right)
                 + \frac{\Delta B_0}{2}A^2
                 + \frac{3v}{4}A^4$, independent of the crystal structure.
                 These terms determine the absolute value of the amplitudes.                 
\item[iii)]
The magnetic coupling term, $\sum_{j=1}^N \aalpha \Ajcc \MMj^2 \Aj$, accounts for the magnetization \mm. It depends on the relative orientation of \kkj and \mm. Thus, as it will be also shown in the following, it breaks the rotational symmetry of the free energy, leads to magnetic anisotropy, and introduces magneto-striction.   
\end{itemize}
Within the long wavelength limit \cite{YHT10}, the approximation of the conserved dynamics for the particle density leads to a non-conserved dynamics for amplitudes. The evolution equations for each amplitude reads
\begin{equation} \label{eq:APFC}
  \frac{\partial \Aj}{\partial t}
  = -|\mathbf{k}_j|^2 \frac{\delta \FF}{\delta \Ajcc},
  \end{equation}
with
\begin{equation}
  \frac{\delta \FF}{\delta \Ajcc} = \left[ B_0^x\mathcal{G}_j^2 +  \aalpha \MMj^2 \right]\Aj + \frac{\partial g^{\rm S}(\sofA)}{\partial \Ajcc}.
\label{eq::APFCevol}
\end{equation}
Thus, the amplitude expansion leads to a 4th order partial differential equation for every amplitude. The number of coupled equations depends on the considered crystal structure, namely on the number of Fourier modes to represent the targeted lattice.

Details on how to generally solve this equations numerically including defects and grain boundaries are reported in Section~4. However, in order to analyze the magneto-structural interaction, we first consider a defect free crystal with constant deformation. This leads to a simpler formulation used to analyse  magnetic-induced anisotropies in the next section.

\section{Symmetry breaking due to magnetization}
\label{sec::symm}
In this section we analyze the proposed magneto-structural coupling in (A)PFC for a single crystal without defects and restrict deformations to be constant in space. In this case, the density, eq.~\eqref{eq::APFCref}, can be expanded according to deformed reciprocal space vectors, $\kkj'=\DDk \kkj$. The deformation matrix $\DDk$ is independent on \kkj and is connected to the spatial deformation of the crystal, $\DD$, by $\DDk=(\DD^{-1})^T$.     
With these assumptions, the free energy, eq.~\eqref{eq::APFCenergy}, reduces to
\begin{equation}
  \FF(\sofA) = |\Omega| \left( \sum_{j=1}^N \Aj \left[B_0^x (1-{\kkj'}^2)^2 - \aalpha (\mm \cdot \kkj')^2 \right] \Aj + g^{\rm S}(\sofA) \right),
  \label{eq::FMAenergy}
\end{equation}
where $|\Omega|$ is the volume of the integration domain. 

\begin{figure}[h]
  \noindent
  \begin{center}
    \includegraphics*[angle = -0, width = 0.55 \textwidth ]{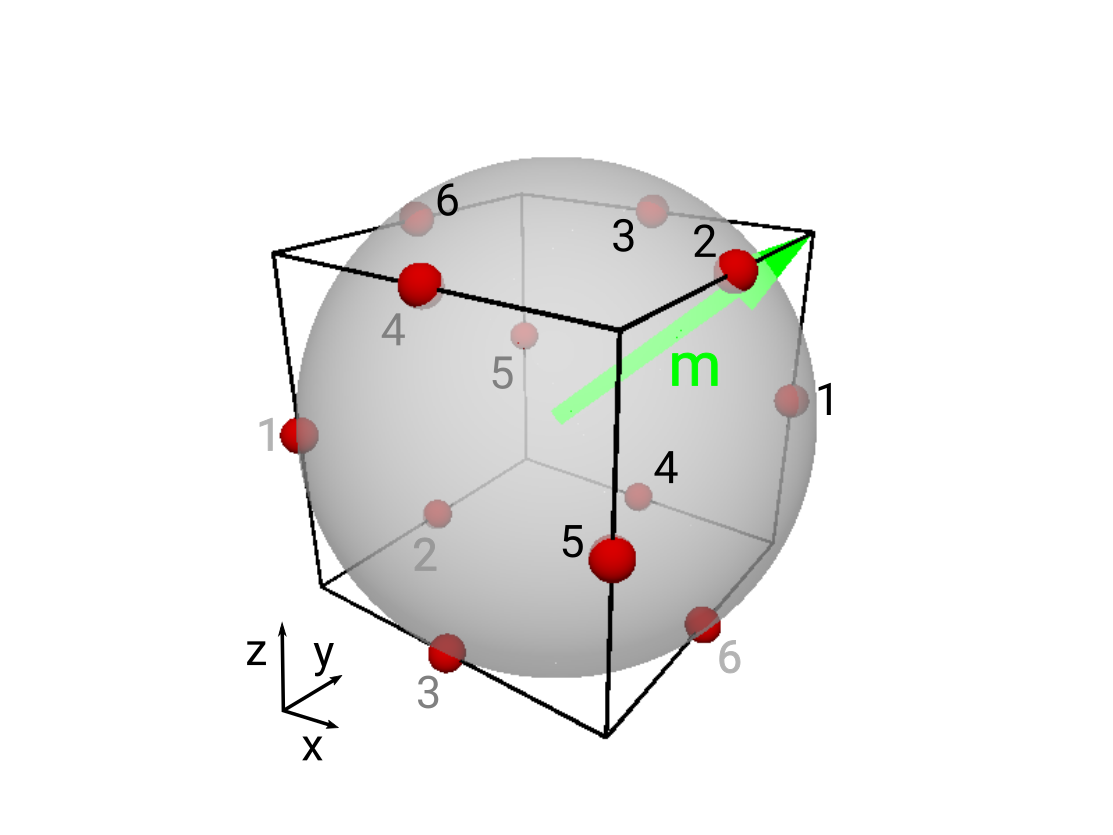}
\begin{minipage}{0.9\textwidth}
\caption{Reciprocal space representation of the BCC lattice. The red spheres represent the vectors $\kkj$ 
and their energetically equivalent vectors $-$\kkj (labelled with the same number in black and grey respectively), representing a BCC crystal in a one mode approximation. Also a specific direction of the magnetization $\mathbf{m}$ discussed in the text is shown.
\label{fig::kjsBCC}
}
\end{minipage}
\end{center}
\end{figure}

We consider a BCC crystal (${\rm S} = \rm{BCC}$), which is described by a set of N=6 \kkj's, see Figure~\ref{fig::kjsBCC}.
The first term, $\sum_j \Aj (B_0^x(1-{\kkj'}^2)^2) \Aj$, is minimized by $|\kkj'| = 1$. It defines the size of the unit cell  of the crystal without any magnetic coupling. The second term, $\sum_j^N \Aj (\mm \cdot \kkj')^2 \Aj$, couples local magnetization to the local structure in the crystal. The structure dependent part of the energy reads  
\begin{equation}
\begin{split}
  f^{\mathrm{BCC}}(\sofA) =& -2\tau(\A_1^*\A_2\A_4+\A_2^*\A_3\A_5+\A_3^*\A_1\A_6 + \A_4^*\A_5^*\A_6^*+{\rm c.c.}) \\
	     & +6v(\A_1\A_3^*\A_4^*\A_5^*+\A_2\A_1^*\A_5^*\A_6^*+\A_3\A_2^*\A_6^*\A_4^*+{\rm c.c.}),  \label{eq::gBCC}
\end{split}
\end{equation}
We recall that this term is independent on deformation. Due to symmetry considerations we only allow for deformations parallel and perpendicular to $\mathbf{m}$. Then the deformation can be written as $\DD=\RR^T \bar{\DD}^{-1} \RR$ with $\bar{\DD}$ a diagonal matrix. $\bar{\DD}_{11}=d_0$ and $\bar{\DD}_{22}=\bar{\DD}_{33}=d_1$ and describes the deformation in direction of $\mm$ ($d_0$) and perpendicular to it ($d_1$). The matrix $\RR$ rotates $\mm$ in the direction of the first eigenvector of $\DD$ with eigenvalue $d_0$. Thus, the deformations are solely defined by expansion or contraction parallel and perpendicular to $\mm$.   

\begin{figure}[htb]
  \noindent
  \begin{tabular}{cc}
      \includegraphics*[width = 0.45\textwidth]{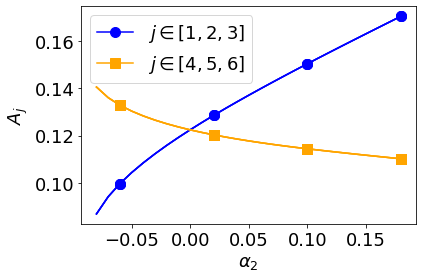}            &
      \includegraphics*[width = 0.46\textwidth]{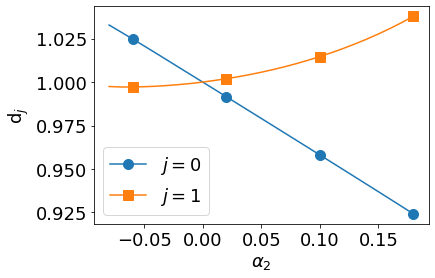}                     
  \end{tabular}
 \begin{center}
\begin{minipage}{0.9\textwidth}
\caption{Symmetry breaking due to magnetization. (left) Amplitudes $A_j$ and (right) deformations $d_0$ and $d_1$ as function of $\aalpha$ for a magnetization $\mathbf{m}$ parallel to $\hkl[111]$. The numbering of \Aj and \kkj follows Figure~\ref{fig::kjsBCC}.   
\label{fig::symBCC}
}
\end{minipage}
\end{center}
\end{figure}

\begin{figure}[htb]
  \noindent
  \begin{tabular}{cc}
    \includegraphics*[width = 0.45 \textwidth]{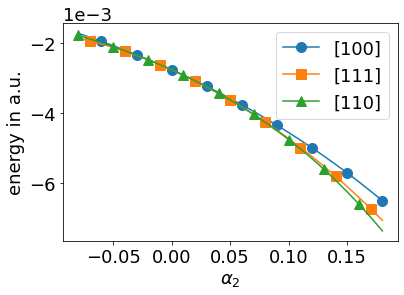} &
    \includegraphics*[width = 0.45 \textwidth]{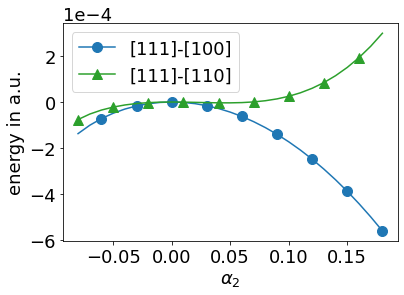} \\
  \end{tabular}
  \begin{center}
\begin{minipage}{0.9\textwidth}
\caption[short figure description]{Anisotropic effect of the magnetization. (left) Total energy and (right) energy difference between different orientations of $\mathbf{m}$ as function of $\aalpha$.  \hkl[100] is always the hard (highest energy) direction of magnetization. The easy (lowest energy) direction is \hkl[110] for $\aalpha< 0$ and \hkl[111] for $\aalpha>0$.
\label{fig::anisoBCC0}
}
\end{minipage}
\end{center}
\end{figure}

We now minimize the energy in eq.~\eqref{eq::FMAenergy} with respect to \sofA and $\DD$ for different orientations of \mm and coupling strength \aalpha. Figure~\ref{fig::symBCC} shows the amplitudes $A_j$ and the deformations $d_0$ and $d_1$ as functions of $\aalpha$ for $\mm$ parallel to the $\hkl[111]$ direction.
The magnetic coupling depends on $(\mm \cdot \kkj)^2$ and thus on the angle of \mm and the reciprocal space vectors describing the crystal, \kkj. Thus, the magnetic coupling acts differently on the single amplitudes. For \mm parallel to \hkl[111], there are two sets of equivalent amplitudes:  those corresponding to $\{\kk_1, \kk_2, \kk_3\}$, which have the same angle with respect to \mm and those corresponding to $\{\kk_4, \kk_5, \kk_6\}$ which are perpendicular to \mm. That is, the magnetic coupling breaks the symmetry of the energy w.r.t. the amplitudes.
This symmetry breaking is shown in Figure~\ref{fig::anisoBCC0}. For vanishing coupling, $\aalpha=0$, all amplitudes are the same. Increasing the coupling and $\aalpha > 0$ leads to increasing amplitudes $\{\A_1, \A_2, \A_3\}$ and decreasing amplitudes $\{\A_4, \A_5, \A_6\}$. The model breaks for \aalpha outside the shown range as some amplitudes vanish, and the energy minimizing state does not describe a BCC crystal anymore. The magnetic coupling for $\aalpha >0$ also leads to a compression of the crystal in the direction of \mm and a smaller contraction perpendicular to \mm.
The total energy decreases with increasing magnetic coupling \aalpha, see Figure~\ref{fig::anisoBCC0}. The magnetic anisotropy induced by the proposed magnetic coupling can be shown by comparing the energy for different orientations of \mm.
The highest energy for a fixed \aalpha is always for \mm parallel to \hkl<100>, therefore referred to as the hard directions of magnetization in this model. The directions with the lowest energy, namely the easy directions, depend on the sign of \aalpha. For $\aalpha > 0$ they are \hkl<110>, while \hkl<111> are the easy directions otherwise.

\begin{figure}[htb]
\noindent
                                                       
\includegraphics*[width=0.95\textwidth] {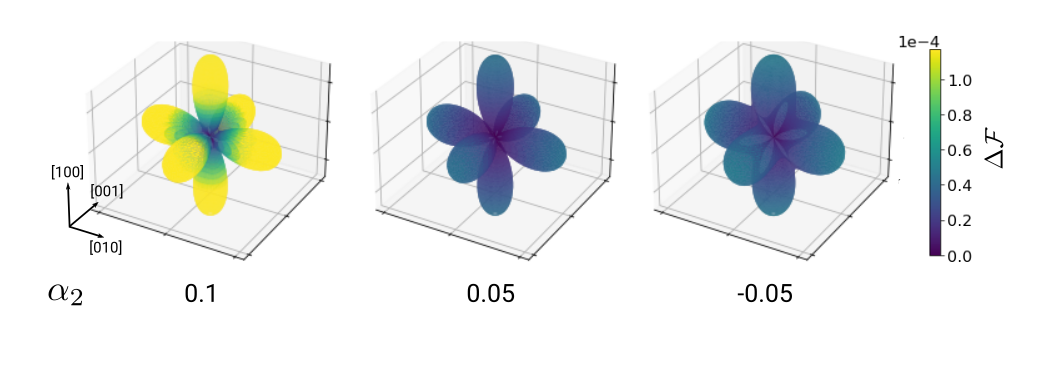} 
\begin{center}
\begin{minipage}{0.9\textwidth}
  \caption[short figure description]{
   Polar representation of energy surface illustrating magnetic anisotropy. The energy difference, $\Delta \FF$, is relative to the minimal energy. The polar angles corresponds to the direction of magnetization, \mm. The radial coordinate is proportional to the normalized energy difference, while the colormap corresponds to $\Delta \FF$. The hard (high energy) direction of magnetization are parallel to \hkl<100> and independent on coupling strength, \aalpha, while the strength of magnetic anisotropy increases with \aalpha. The symmetry of the magnetic anisotropy resembles the cubic symmetry of the crystal. 
\label{fig::scanBCC}
}
\end{minipage}
\end{center}
\end{figure}

The general dependency of the energy can be directly explained by the sign of the magnetic coupling. For $\aalpha < 0 $ the magnetic coupling term introduces a positive contribution in the energy and, thus, the total energy increases. However, the magnetic anisotropy depends on the relative influence of the magnetic coupling on the different amplitudes.
In order to understand the induced magneto-striction, the influence of the magnetic coupling on \sofkdef has to be examined. The term $(1-(\kkj')^2)^2$ defines the $\kkj'$ in equilibrium without magnetic interaction. That is, this term has a minimum at $|\kkj'|=1$. The energy contribution of this term vanishes in this case.
The magnetic coupling changes the situation. The minimum of $(1-(\kkj')^2)^2-\aalpha (\mm\cdot \kkj')^2$ depends on the relative orientation of $\kkj'$ and \mm. For $\aalpha >0$, the minimum is shifted towards larger $\kkj'$, and the energy contribution becomes negative. Thus, amplitudes influenced by this term, namely $A_j$ for which $\mm \cdot \kkj' \neq 0$, become larger. The shift of the minimum towards larger $\kkj'$ leads to a compression of the crystal in this direction. This analysis is exact for a one-dimensional crystal (so namely for a single \kkj vector). For a structure as BCC it is more involved as the \kkj's do not enter in the energy independently. They all can only transform with the same deformation, $\DDk$.   
Extending these computations to all orientations allows to compute the magnetic anisotropy. The corresponding polar plots for various $\aalpha$ are provided in Figure~\ref{fig::scanBCC}.

\section{Numerical approach and setup}
\label{sec::num}
\begin{figure}[htb]
  \noindent
  \begin{center}
  \begin{tabular}{cccc}
    &\includegraphics*[angle = -0, width = 0.35 \textwidth ]{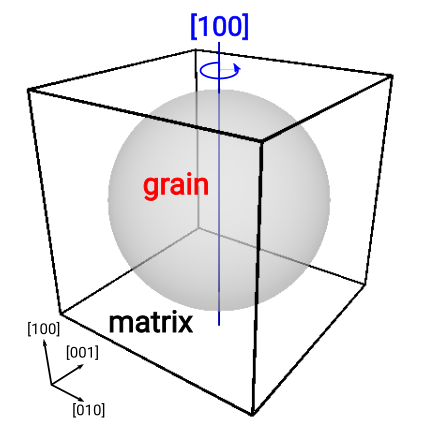} &
                                                                                 &\includegraphics*[angle = -0, width = 0.35 \textwidth ]{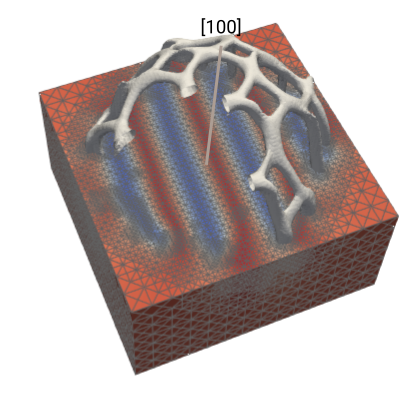}                                                                              
    \end{tabular}
\begin{minipage}{0.9\textwidth}
\caption[short figure description]{
  Setup and APFC representation for the simulation of grain shrinkage. (left) Schematics of the the single relaxed crystal (matrix) embedding a spherical grain rotated by $\arot=10 \degree$ about the \hkl[100] axis. (right) APFC representation of the rotated grain in the relaxed matrix. The color shows ${\rm Re}(\A_0)$. The dislocation network is visualized in gray as the isosurface $\overline{\Asq} = 0.073$. In addition, the spatial adaptivity of the FEM implementation, featuring different levels of refinement at the matrix, the rotated grain, and the defects is shown.
 \label{fig::APFCsetUp} }
\end{minipage}
\end{center}
\end{figure}

We solve eqs.~\eqref{eq:APFC} numerically using the finite element method (FEM). At variance with other common approaches exploited in the literature, such as the pseudo-spectral Fourier method, it allows for an optimized spatial discretization for amplitudes that vary unevenly in space. Indeed, they feature significant variations at dislocations, requiring a resolution approaching the one needed by the classical PFC model, but vary slowly for rotated and strained crystals and are constant for relaxed crystalline domains. We consider in particular the approach reported in references \cite{SBE17,PSV19} and adapt it to the novel coupling with magnetic fields proposed in this work. To discretize eqs.~\eqref{eq:APFC} we expand the fourth-order operator 
\begin{align}
  \left[B_0^x \GGj^2+ \aalpha \MM_j^2 \right]
  =(\sqrt{B_0^x} \GGj+ \ii \sqrt{\aalpha}\MMj)(\sqrt{B_0^x}\GGj- \ii \sqrt{\aalpha}\MMj)
  =:\NNjp \NNjm,
\end{align}
and rewrite eqs. \eqref{eq:APFC} as systems of second order equations
\begin{equation}
\begin{split}
  \frac{\partial \Aj}{\partial t} &=  -|\mathbf{k}_j|^2 \left[ B^x_0 \NNjp \mu_j  + G_j(\sofA) \right], \\
  \mu_j&= \NNjm \Aj,
\end{split}
\end{equation} 
with $G_j(\sofA) := {\partial g^{\rm S}(\sofA)}/{\partial \Ajcc}$ the nonlinear terms. We consider a semi-implicit time discretization evaluating linear terms implicitly and providing an approximation for the implicit evaluation of non linear terms. In particular, these terms are linearized for every amplitude via a one-iteration Newton method, such that  
\begin{equation}
\begin{split}
  G_j(\{\Aj^{n+1}\}) &\approx G_j(\{\Aj^n\})+(\Aj^{n+1}-\Aj^n)  \left. \frac{\partial G_j(\sofA)}{\partial A_j} \right|_{\sofA=\{\Aj^n\}} \\ 
                         &=: G_j^{\mathrm ex}(\{\Aj^n\}) + G_j^{\mathrm im}(\{\Aj^n\}) \Aj^{n+1},
\end{split}
\end{equation} 
where $\Aj^n$ and $\Aj^{n+1}$ denote $\Aj$ at timestep $n$ and $n+1$, respectively. The considered semi-implicit discretisation then reads
\begin{align}
  \begin{bmatrix}
    1+\Delta t  |\mathbf{k}_j|^2G_j^{\mathrm im}(\{\Aj^n\})  & \Delta t |\mathbf{k}_j|^2 \NNjp  \\
    - \NNjm & 1
  \end{bmatrix}
  \begin{bmatrix}
    \Aj^{n+1} \\
    \mu_j^{n+1}
  \end{bmatrix}
  =
  \begin{bmatrix}
    \Aj- \Delta t |\mathbf{k}_j|^2 G_j^{\mathrm ex}(\{\Aj^n\}) \\
    0
  \end{bmatrix},
\end{align}
where $\Delta t$ is the timestep. For the investigations reported in this work, a proper numerical convergence is achieved for $\Delta t=1$. In order to use standard solvers and real basis functions for the FEM implementation, the above system of equations has to be split further into the real and imaginary parts of the amplitudes. This leads to a system of four coupled second-order equations for each amplitude which can be solved by exploiting linear elements. See references \cite{SBE17,PSV19} for further details. The weak form of these systems of equations and their FEM discretization are implemented in the parallel and adaptive finite element toolbox AMDiS \cite{Vey2007,Witkowski2015}. Following reference~\cite{PSV19} mesh refinement is set according to the local variation of the amplitudes. In particular, the spatially-dependent wavelength of their oscillation is computed and properly resolved by a fixed number of mesh elements per period of the oscillation. In addition, a minimum refinement is set for regions where the crystal does not exhibit any deformation or rotation, and a fine mesh is set at dislocations (see also figure~\ref{fig::APFCsetUp}).


We examine the influence of magnetization on the shrinkage of an initially spherical grain rotated with respect to the surrounding matrix, see figure~\ref{fig::APFCsetUp}, in a BCC crystal. The grain has a radius of 30$\pi$ and is rotated by $10\degree$ about the \hkl[100] direction. That is, the rotation axis is set as the z-axis in our frame. From a macroscopic point of view, the normal velocity of the grain boundary $v$ can be described by \cite{Mullins_JAP_1956,DHH1997}
\begin{align}
 v = -M (\gamma \kappa - \Delta f),
 \label{eq:mcf}
\end{align}
where $M$ is the mobility, $\gamma$ is the grain boundary energy, $\kappa$ the local curvature and $\Delta f$ the difference of the bulk energies of matrix and grain, and neglecting additional elasticity effects \cite{Zhang2017,HAN2021}.
In our case $\Delta f$ results from the magnetization, as the magnetic properties are different in the grain and the matrix. 
Without $\Delta f$, eq.~\eqref{eq:mcf} reduces to mean curvature flow and describes isotropic grain shrinkage due to minimization of the grain boundary energy. As already shown by PFC \cite{Yamanaka2017} and APFC \cite{Salvalaglio2018} simulations without magnetization, while recovering the general scaling described by eq.~\eqref{eq:mcf}, the shrinkage is anisotropic. We will therefore 
directly compare our result for the grain shrinkage under the influence of magnetization with these previous findings.

The crystal structure in the matrix surrounding the circular embedded grain is set as the reference structure. Thus, the amplitudes therein are constant and real. Within the uniformly rotated grain, the amplitudes vary regularly, still featuring a constant $\Asq$. 
This quantity varies significantly at dislocations only, and can be therefore exploited for their identification \cite{SBE17,Salvalaglio2018}. Indeed, we recall that dislocations are described in the APFC model as zeros of some of the complex amplitudes, namely the ones which have singular phases \cite{Salvalaglio2022}. Far away from the dislocation core, they smoothly connect to their bulk values. A diffuse representation of the dislocation core over a few lattice spacing is then achieved \cite{SalvalaglioJMPS2020}, consistent with non-singular continuous descriptions \cite{Lazar2005,Cai2006}. Dislocation networks can then be reconstructed as regions where $\Asq<\overline{\Asq}$ with $\overline{\Asq}$ a selected threshold. For visualization purposes, the extension of these regions across the dislocations can be chosen arbitrarily by selecting a value of $\overline{\Asq}$ within the range of $\Asq$, provided that it ensures enough resolution \cite{SBE17,Salvalaglio2018}. Here, we set $\overline{\Asq} = 0.073$., corresponding to $\sim 2$ lattice spacing.
Figure~\ref{fig::APFCsetUp} illustrates the setup for the simulation of the embedded rotated grain and the resulting small-angle grain boundary, together with the features of the amplitudes, the reconstructed dislocation network after relaxation of the initial condition, and the inhomogeneous mesh refinement used for simulations.

\section{Simulation results}
To study the influence of the magnetization on grain shrinkage, three configurations are considered: a simulation without magnetization, $\aalpha =0$, (M0), corresponding to \cite{Yamanaka2017,Salvalaglio2018}, and two settings with magnetization, one that hinders grain shrinkage (M1) and one that enhances grain shrinkage (M2). We consider a coupling strength, $\aalpha = 0.1$, and only vary the direction of magnetization, \mm, which is chosen according to the magnetic anisotropy to enhance or hinder grain boundary motion.  
\begin{figure}[htb]
  \noindent
  \begin{center}
  \begin{tabular}{cccc}
    &\includegraphics*[angle = -0, width = 0.4 \textwidth ]{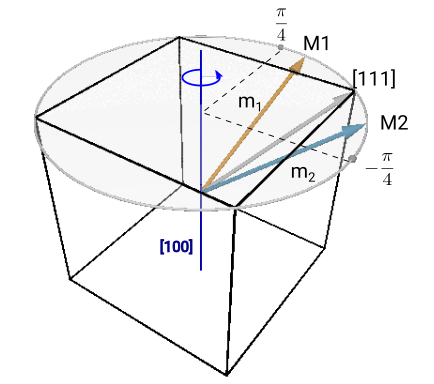} &
    &\includegraphics*[angle = -0, width = 0.45 \textwidth ]{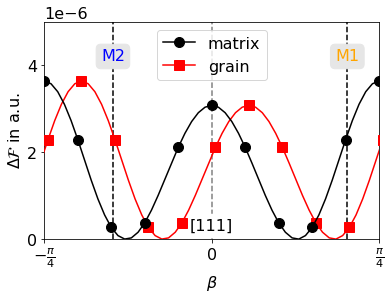}                                                                              
    \end{tabular}
\begin{minipage}{0.9\textwidth}
  \caption[short figure description]{Definition of the magnetization for the computational investigations. (left) Problem setup for (M1) and (M2). The corresponding magnetizations $\mm_1$ and $\mm_2$ are chosen to highlight differences due to the magnetic anisotropy of the matrix and the grain. (right) Energy difference for rotation $\beta \in [-\frac{\pi}{4}, \frac{\pi}{4}]$ about $\hkl[111]$.  The two cases (M1) and (M2) are explicitly marked. 
 \label{fig::setUpMag} }
\end{minipage}
\end{center}
\end{figure}
For $\mm$ parallel to $\hkl[111]$, the energy in the matrix would be larger than in the grain. This results from the grain rotation by $\arot$ about \hkl[100] of the matrix crystal. The specific setup can be seen in Figure~\ref{fig::APFCsetUp}. This energy difference would allow us to consider the effect of the anisotropy induced by the magnetization. However, as the differences in the magnetic anisotropy for this configuration are relatively low (see also Figure~\ref{fig::setUpMag}), we consider a different orientation for \mm. We rotate \mm about the \hkl[100] direction by $\beta={\pi}/{8}+{\arot}$ for (M1) and $\beta=-{\pi}/8-{\arot}/{2}$ for (M2), respectively.
For positive $\beta$ (M1), the bulk energy in the grain is lower than in the matrix. 
Thus, $\Delta f$ in eq.~\eqref{eq:mcf} becomes positive, which reduces the velocity of the grain boundary. Rotation of \mm in the other direction (M2) leads to lower bulk energy in the matrix than in the grain. Thus, $\Delta f$ becomes negative, which provides an additional driving force in the direction of grain shrinkage and thus enhances grain shrinkage.  
 
\begin{figure}[htb]
  \noindent
  \begin{center}
  \begin{tabular}{cc}
    \includegraphics*[angle = -0, width = 0.45 \textwidth ]{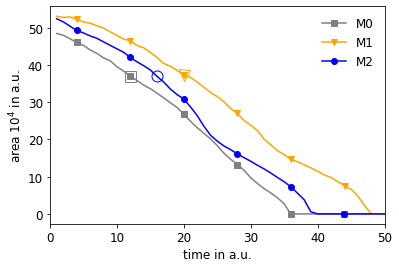} &
    \includegraphics*[angle = -0, width = 0.455 \textwidth ]{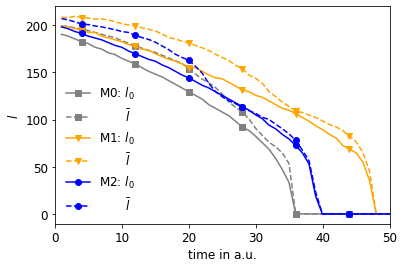}
    \end{tabular}
\begin{minipage}{0.9\textwidth}
\caption[short figure description]{
Grain size vs. time. (left) The area of the grain boundary decreases over time. (M0) corresponds to shrinkage without magnetization (\aalpha=0). For (M1), the grain has lower energy than the matrix due to magnetization, while the opposite configuration is realized for (M2). (right) The grain shape is represented by its diameter in \hkl[100] direction, $l_0$, and its mean diameter in \hkl(100) plane, $\bar{l}=\frac{1}{2}(l_1+l_2)$, see also Figure~\ref{fig:shapeComp}.   
 \label{fig:areaEvol}
}
\end{minipage}
\end{center}
\end{figure}
\begin{figure}[htb]
  \noindent
  \begin{center}
    \includegraphics*[angle = -0, width = 0.75 \textwidth ]{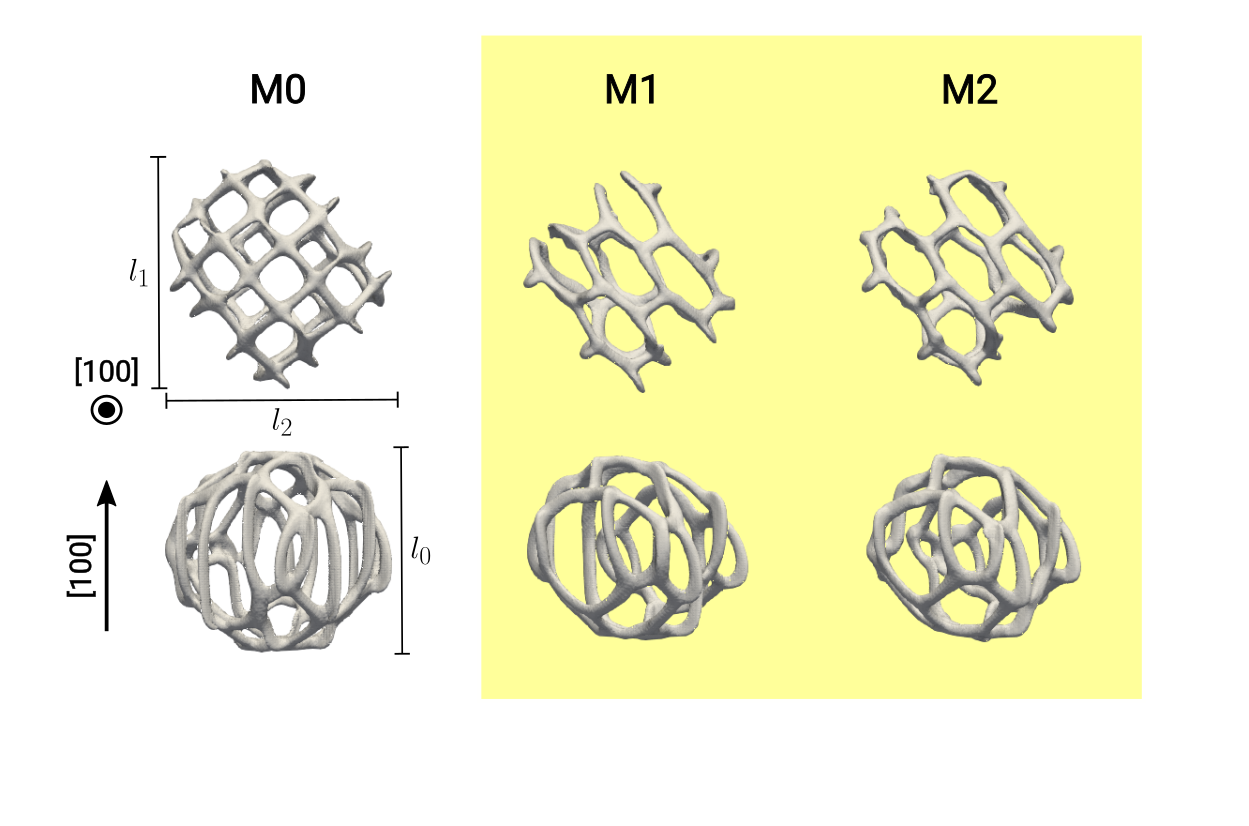}
\begin{minipage}{0.9\textwidth}
\caption[short figure description]{
Dislocation networks for the three cases (M0), (M1) and (M2) for similar grain size. Different orientations are shown. These configurations correspond to the open symbols in Figure~\ref{fig:areaEvol}(left). $l_0$, $l_1$ and $l_2$ used in Figure~\ref{fig:areaEvol} are also illustrated on the dislocation networks obtained for (M0).
 \label{fig:shapeComp}
}
\end{minipage}
\end{center}
\end{figure}

Figure~\ref{fig:areaEvol} shows the shrinkage of the grain for the three cases (M0), (M1), and (M2). We consider the area of the grain boundary, computed as in \cite{Salvalaglio2018}, as well as lengths $l_0$ and $\bar{l}=\frac{1}{2}(l_1+l_2)$, as defined in Figure~\ref{fig:shapeComp}, to account for the anisotropic shrinkage. As expected in the setting of enhanced grain shrinkage (M2), the grain shrinks faster than in the case of hindered grain shrinkage (M1). However, the fastest shrinkage is obtained for the reference case (M0). This behavior has already been observed in \cite{BEV19}, and it can be explained by the dependency of $M\gamma$, namely the product of the mobility and the grain boundary energy, on \mm. The anisotropy of the shrinkage is influenced by $\mm$ but remains within the same order for all three cases.

To further highlight details of the grain shrinkage discussed above, dislocation networks corresponding to the open symbols in Figure~\ref{fig:areaEvol}(left) are shown in Figure~\ref{fig:shapeComp}. They correspond to the same grain boundary area and indicate the deformation of the dislocation network by the magnetization.

\begin{figure}[htb]
  \noindent
  \begin{center}
    \includegraphics*[angle = -0, width = 0.75 \textwidth ]{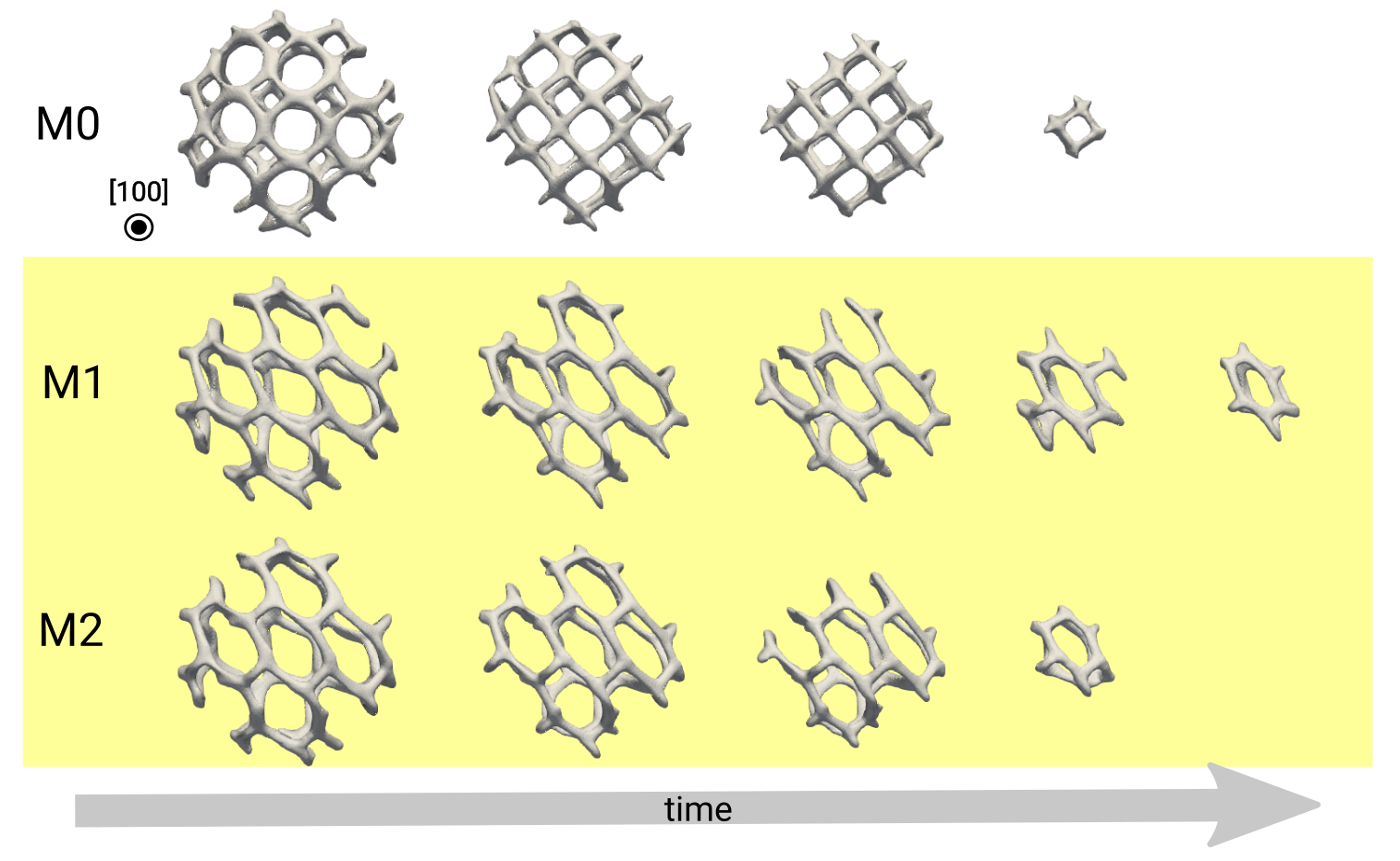}
\begin{minipage}{0.9\textwidth}
\caption[short figure description]{
Evolution of the dislocation networks for the three cases (M0), (M1), and (M2) during grain shrinkage. The snapshots correspond to states represented by the full symbols in Figure~\ref{fig:areaEvol}.
 \label{fig:shapeEvol}
}
\end{minipage}
\end{center}
\end{figure}

Figure~\ref{fig:shapeEvol} shows the dislocation networks during their shrinkage over time. Representative stages are selected, corresponding to the full symbols in Figure~\ref{fig:areaEvol}. They also indicate the magnetization-dependent deformation of the dislocation network. Besides the global shapes, the dislocation networks also show a qualitatively different arrangement of defects for different magnetizations.
\label{sec::res}
\section{Discussion}

The classical classical APFC model has been proved powerful in describing crystalline structure and defect networks at coarse-grained and mesoscale lengthscales while exploring diffusive timescale similarly to the PFC model.
In \cite{Salvalaglio2018} it has been shown that the dislocation network and its (overdamped) dynamics predicted by APFC corresponds to the one achieved by PFC \cite{Yamanaka2017}. More importantly, static properties, symmetries and energetics, are qualitatively as expected by other atomistic models \cite{Yang2010} and experiments \cite{Schober1969}. It is worth mentioning that, although minimal formulation have been mostly considered, PFC and APFC formulations may be adapted to fit quantitatively grain boundary properties of specific materials \cite{JAA10,Hirvonen2016}.
In this work, we showed that the influence of strong magnetic fields on the dynamics of grain boundaries could be included in the APFC model. Depending on the orientation of magnetization, the grain growth is hindered or enhanced as predicted by continuum models, eq.~\eqref{eq:mcf}. The dislocation network is deformed due to the deformation of the crystal and matrix due to magneto-striction. Although the qualitative character of the reported results, extended parameterizations may be devised and extended model may be considered together with the coupling with magnetic field here proposed.

Many relevant magnetic materials are alloys where material properties are dependent on the solute concentration. This give rise to variation of magnetic properties due to variation in solute concentration. Especially at grain boundaries segregation may lead to change of magnetic properties \cite{Kir07,Kir07b,STT09} and, vice-versa, segregation is influenced by magnetization \cite{WK21,IYN85}. 
PFC and APFC models for binary alloys and ordered crystals have been already reported in literature \cite{EHP10,Elder2007,SP17,AEV17,TDH19,SalvalaglioPRL2021}. It has been demonstrated that these models can track grain boundary segregation \cite{SP14} and precipitation \cite{SP17}, as well as dislocation dynamics accounting for Cottrell atmospheres \cite{SalvalaglioPRL2021}. These models are formulated either by treating the density of every species as a density field or by reformulating the model in terms of a mean mass density and a concentration field. However, in both approaches, the magneto-structural interaction can easily be in-cooperated the same way as proposed in eq.~\eqref{eq::cntrPFCenergy}. The magnetization is coupled to the gradient of density the density field, $\aalpha \phii (m \nabla)^2 \phii$, leading to magnetostriction and magnetic anisotropy if \phii is a density field representing a crystal. Also, the coupling constant \aalpha may be set as dependent on local concentration to account for magneto-structural interaction due to solute concentration at dislocations and grain boundaries. Additional terms which are isotropic but depend on magnetization and concentration may be introduced to account for local variations in the crystal, e.g., $\mm^2 \phii^2$ \cite{FPK13,SSP15}. However, it is worth noting that this ansatz is not expected to be sufficient for an accurate, quantitative description of any crystalline material. Indeed, magnetic anisotropy and magnetostriction are controlled by a single parameter \aalpha, and, for BCC, the hard direction in this model is always \hkl[111]. Thus, higher-order terms for comprehensive modeling of magnetic coupling are needed as proposed in \cite{SSP15}. Further research is ongoing to develop a minimal magnetic coupling model that can be adapted to a broader class of materials within the APFC model.       
Furthermore, to account for energies due to variation in magnetization, a formulation beyond the assumption of constant magnetization \cite{BV20} is necessary. To this goal, the free energy has then to be complemented by a Landau-Lifshitz-Ginzburg type energy for the magnetization field as done in \cite{FPK13,SSP15}.  

In most of the extensions and perspectives mentioned above, the magneto-structural properties in the crystal can still be explained by an analysis similar to what has been reported in this work. In particular, the simplest magneto-structural coupling as studied here is the basic building block to introduce anisotropic magnetic interaction in the modeling of binary materials with PFC and, thus, APFC. This paves the way for the development of APFC models including magneto-structural interaction for alloys and, therefore, towards quantitative application of the APFC model to the modeling of specific materials.

\section{Conclusion}
\label{sec::con}

We have given a proof of concept for the simulation of the interactions between magnetic fields and dislocation networks in a three-dimensional setting with the APFC model. We derived a magnetic APFC model and adapted an existing, state-of-the-art numerical approach to solve the corresponding equations. As a result, the basic physics of magneto-structural interactions are accounted for in a multiscale approach.
We have demonstrated that the considered framework accounts for magnetic anisotropy and provided an example of its influence on the shrinkage of a rotated spherical grain within an unrotated matrix. This consists of a minimal setting that may be extended toward the modeling and simulation of defect networks and microstructure evolution on diffusive time scales and at relatively large length scales. The proposed approach is indeed the basic building block to introduce magneto-structural coupling in PFC models of complex crystals and alloys.
However, various open problems remain. One central issue is an appropriate parametrization of the material parameters for specific materials. While this has been done for the bulk modulus and the grain boundary energy, e.g., for Fe \cite{Jaatinen2009}, the magnetic anisotropy in the model has to be tuned to resemble known functional forms for the magnetic anisotropy of materials of interest.

\ack
AV and RB acknowledge support by the German Research Foundation (DFG) within SPP1959 under Grant No. VO899/20-2. MS acknowledges support from the Emmy Noether Programme of the German Research Foundation (DFG) under Grant No. SA4032/2-1. We further acknowledge computing resources provided at J\"ulich Supercomputing Center under grant PFAMDIS.

\section*{References}
\bibliographystyle{iopart-num-mod.bst}
\bibliography{collectAll}

\end{document}